\documentclass[12pt]{article}

\def\beq{\begin{equation}}
\def\eeq{\end{equation}}
\def\bea{\begin{eqnarray}}
\def\eea{\end{eqnarray}}
\def\C{C}

\def\ym{y^{-1}}

\def\l{\lambda}
\def\o{\omega}
\def\vm{v^{-1}}
\def\D{D^{\omega }(G)}
\newcommand{\elt}[2]
{ {\scriptstyle #1} | \! \raisebox{-.2ex}{\underline{\makebox[0.6em]{}}}
   \hspace{-0.6em}
   \raisebox{-2ex}{$\,\scriptstyle #2 \,$} }
\begin{document} 

{\LARGE Representation Theory of Twisted} \\ 
\vskip .1cm 
\hskip 3cm {\LARGE Group  Double } \\
\vskip .3cm 
 {\large  D. Altschuler\footnote{4 rue Beausite,CH  1203 Gen{\`e}ve,
 Suisse} }, 
 {\large A. Coste \footnote{LPT CNRS UMR 8627, batiment
 210-211, Universit\'e d'Orsay,  F 91405 Orsay Cedex}  } 
and 
{\large J-M. Maillard\footnote{ LPTHE, Tour 25, case 7109, 2
 Place Jussieu, 75251 Paris Cedex 05}~\footnote{ 
 e-mail: maillard@lpthe.jussieu.fr}
  }

\date{\today}

\vskip .8cm 

\vskip 1.cm 
\begin{abstract}
 This text collects useful results concerning  
the quasi-Hopf algebra $\D  $.  
We give a review of issues  related to its use in conformal theories and 
 physical mathematics. Existence of such algebras based
 on 3-cocycles with values 
 in $  {R} / {Z}  $ 
 which mimic for finite groups  Chern-Simons terms  
of gauge theories, open 
 wide perspectives in the so called "classification program".
  The modularisation 
  theorem proved for quasi-Hopf algebras by  two authors some years ago 
  makes the computation of topological invariants possible.  
  An updated, although partial,  bibliography of recent developments is
  provided. 

\end{abstract}
\vskip .1cm 
\vskip .1cm 

\vskip .8cm 

 \noindent
\vskip .2cm 

\vskip .8cm 
\vskip .8cm 
\pagebreak 

{\large \bf Introduction }
\vskip .2cm 
Lattice statistical models 
with boundary conditions or quantum field theories 
with boundary conditions, have been extensively studied in the last 
ten years. 
In an integrable framework and from an algebraic point of view, 
different boundary conditions may lead to various interesting
algebras. 

Because of dynamical interactions between excitations, a phenomenon
sometimes called ``fusion of excitations 
together'' takes place as a consequence of
a non trivial geometric transformation. This idea 
lies at the heart of the 
celebrated Frobenius-Pasquier-Verlinde formula~\cite{VerlindeHisto}. 
Later, R. Dijkgraaf, C. Vafa, E. Verlinde and H. Verlinde 
applied this to
orbifolds.   

 
Such a fusion corresponds, in the mathematical 
formalism, to a coproduct
morphism being part of quasi-Hopf (or some quantum algebra) axioms. 

Such a geometric transform can be a $\, SL_n(Z)\, $ diffeomorphism of
the $\, n$-torus, or a ``tour'', along a non-trivial homological
cycle. Note that more complicated manifolds are also a very 
fascinating field of research. 

In two-dimensional integrable systems compatibility between the
Yang-Baxter equations and 
the fusion is an interesting general issue. 

To our knowledge, R. Dijkgraaf and E. Witten pioneered the use of 
group 3-cocycles as a
discrete analogue of flat connections on 
3-manifolds~\cite{witten}. 
This viewpoint
makes very clear the role of local gauge invariance in building up
topological invariants. However, their
 paper remains in the framework
of so-called ``topological theories'' and cobordism, which 
 can be fruitfully associated with a 
more algebraic, and algorithmic,
 approach such as the one initiated by V. G. Drinfeld,
M. Jimbo, T. Miwa, N. Reshetikhin, V. Turaev,
and many others~\cite{drinfeld, reshet}.
This is the viewpoint we will detail here. In some conference
proceedings~\cite{dpr},
 V. Pasquier et al. 
obtained a nice breakthrough when  formulating
a twisted double as a quasi-Hopf algebra, called $\D $. 
This is mostly this work that we revisit in 
this article\footnote{which is a developed version of the preprint 
 IHES/M/99/52}, including new material and more general 
approach of classification problems.

In a series of papers
D. Altschuler and A. Coste proved that N. Reshetikhin and V. Turaev
functorial construction
extends to quasi-Hopf algebras and therefore, the invariants from 
R. Dijkgraaf and E. Witten topological 
theories have been conjectured by 
them to be exactly recovered from a link-surgery Markov 
trace~\cite{witten}.

It is clear considering simple lens spaces that this equivalence goes
between triangulation and Dehn surgery of the same 3-manifold. The
proof of these equalities, as well as consideration of Heegaard
splitting presentation, has been provided to our knowledge
by S. Piunikhin and followers~\cite{piunikhin}.

\vskip .5cm 
{\large \bf A few heuristic recalls on Drinfeld's Quantum Double}
\vskip .2cm 

The interest of quasi-triangular Hopf algebras is that they produce 
naturally solutions of the Yang-Baxter equations in a ``universal form'', 
in a first approach,   
{\em without spectral parameter}. One can build
quasi-triangular Hopf algebras starting from any Hopf algebra 
introducing two Hopf algebras dual to each other (see for 
instance~\cite{Nill,Novikov}).
 Two Hopf algebras 
(isomorphic as vector space) are dual to each other if the product and
co-product of the first Hopf algebra coincide with the co-product 
and product of the other one. Let  $\, e_a\, $ be the basis vectors
 of the first Hopf algebra, and  $\, e^{a}\, $ the basis vectors
 of the second Hopf algebra, we require :
\begin{eqnarray}
&&e_a \, e_b \, = \, \, \, C^{c}_{a,\,b} \cdot  e_c\, ,  \qquad 
\Delta(e^{a}) \, = \, \, \, C^{a}_{c,\,b} \cdot e^{b} \otimes e^{c}
             \nonumber \\  
&&           \nonumber \\   
&&e^{a} \, e^{b} \, = \, \, \, \tilde{C}_{c}^{a,\,b} \cdot  e^{c}\, ,  \qquad 
\Delta(e_{a}) \, = \, \, \, \tilde{C}_{a}^{b,\,c}  \cdot e_{b} \otimes e_{c}
\end{eqnarray}
where the constants of structure $\, C^{c}_{a,\,b}\, $ 
and $\, \tilde{C}_{c}^{a,\,b}\, $ are such that the two Hopf algebras
are co-algebras. The antipodes  of the two Hopf algebras, $\, \gamma$ 
and $\, \rho$  :
\begin{eqnarray}
\gamma(e_a) \, = \, \,\gamma_a^{b} \cdot  e_b\, , \qquad \qquad 
\rho(e^a) \, = \, \,\rho^a_{b} \cdot  e_b
\nonumber 
\end{eqnarray}
are related by a simple matrix
 inversion : $\, \rho \, =\, \, \gamma ^{-1}$.

One introduces a normal ordering when assembling together
the two Hopf algebras (think for instance that one Hopf algebra
 corresponds to creation operators, and the other
to annihilation operators).  
%
%
The coproduct of this double Hopf algebra can be defined as follows :
\begin{eqnarray}
\Delta(e_a \, e^{b}) \, = \, \, \tilde{C}_{a}^{c,\,d} \cdot 
 C^{b}_{e,\,f} \cdot 
e_c \, e^{f} \otimes e_d \, e^{e}
\end{eqnarray}
The transposed coproduct in the double is just :
\begin{eqnarray}
\Delta'(e_a \, e^{b}) \, = \, \, \tilde{C}_{a}^{c,\,d} \cdot 
 C^{b}_{e,\,f} \cdot 
e_d \, e^{e} \otimes e_c \, e^{f} 
\end{eqnarray}
The coproduct and the transposed coproduct 
are intertwined by the following $\, R$-matrix :
\begin{eqnarray}
R \, \Delta (x) \, = \, \, \Delta'(x) \, R\, , \qquad \hbox{where :}
\qquad 
R \, = \, \, \sum_a \, e_a \otimes {\bf 1} \otimes  {\bf 1} 
                           \otimes       e^{a}
\nonumber
\end{eqnarray}
The antipode in the double is an antihomomorphism that
we will not write here. The prescription to get the normal ordering is 
that the product and coproduct are compatible.
 Drinfeld's normal
ordering is :
\begin{eqnarray}
  :\  e^{a} \, e_b \  :\,\, \, = \, \,\, \,
C^{e}_{f,\,d} \cdot 
C^{a}_{e,\,g} \cdot \tilde{C}^{h,i}_{b} \cdot  \tilde{C}^{c,f}_{i} 
              \cdot \rho_h^{g}          \cdot e_c \, e^{d}
\nonumber 
\end{eqnarray}
where we use the Einstein's conventions for summation
over up and down indices (contraction).  

\vskip .5cm 
{\large \bf Identification of the representations of $\D $ }
\vskip .2cm 
\par In two very dense papers V.G. Drinfeld \cite{drinfeld} has 
developed the notion of quasi-triangular quasi-Hopf algebra, which
seems particularly fruitful in 
Geometry and Physics~\cite{witten,reshet}.
Examples  of such a structure can be obtained as follows:  
\par Let $G$ be a finite group with unit $e$ and order $|G|$ . 
 Let ${\o}$ be a normalized 3-cocycle with 
values in $U(1)$, i.e. ${\o}$ is an 
application from $G^3$ to $U(1)$ satisfying for any 
$(g_1 , g_2 , g_3 , g_4 ) \in G^4$ :
\beq {\o}(g_1,g_2,g_3) \  {\o}(g_1,g_2 g_3,g_4) \  {\o}(g_2,g_3,g_4) = 
                       {\o}(g_1 g_2,g_3,g_4) \ {\o}(g_1,g_2,g_3 g_4)         
\eeq 
\beq {\o}(g_1,g_2,e) = {\o}(g_1,e,g_2) =     {\o}(e,g_1,g_2)   
\eeq 
Where $e$ is the unit element of $G$. In this multiplicative 
notation $\omega $ is the exponential of an additive cocycle 
with values in $ i\ R / 2 \pi Z  $.   
These equations imply a number of identities
 which are derived in the appendix.
\vskip .3cm 
\par From these data one defines~\cite{dpr} a finite 
dimensional algebra over the 
complex numbers as the linear span of the $|G|^2$ generators
$ ( \elt{g}{x} )_{(g,x)\in G^2 } $ satisfying the relations 
$ \elt{g_1}{x} \cdot \elt{g_2}{y} 
 = \delta_{g_1,\ xg_2 x^{-1}}\  \theta_{g_1}(x,y) $  
\beq \hbox{where} \ \ \ \theta_{g}(x,y) = 
   {\o}(g,x,y)\  \omega (x,y,(xy)^{-1}gxy) \ {\o}(x,x^{-1}gx,y)^{-1}. 
 \eeq
$\D$ is associative by virtue of (\ref{a1}) and has unit element
$ 1_{\D} = \sum_{g\in G} \elt{g}{e} $. It can be equipped with a ribbon 
quasi-Hopf quasi-triangular structure described
 in details in~\cite{dpr,ac} allowing one to compute 
topological invariants of links 
 and three manifolds by the method explained in \cite{reshet,ac}. Here 
we wish to present a study of the representations of $\D$ which are the 
objects "colouring" the links in this method.  

\vskip .5cm 
{\bf Proposition 1}: Any finite dimensional representation of 
$\D $ is a direct sum of irreducible ones.
\vskip .1cm 
Proof: Let $(\pi,V)$ be a representation with a non trivial stable 
subspace $W$. Let $P$ be any projector onto $W \ (P^2=P, $  $Im P=W, $
$KerP \bigoplus W = V )$. Define
$$ P_o = \frac{1}{|G|}\sum_{(g,x)\in G^2}
         \frac{1}{\theta_{g^{-1}}(x,x^{-1}) }
\pi( \elt{x^{-1}g^{-1}x }{x^{-1} } ) P \pi (\elt{g^{-1}}{x}  )\ $$
This formula implies $Im P_o \subset Im P $. Furthermore, because of
(\ref{a1}) derived in the appendix below,   
$\theta_{x^{-1}g^{-1}x }(x^{-1},x)=\theta_{g_{-1}}(x,x^{-1})$
so that ${P_o}_{|Im P}= Id_{Im P}$, implying $Im P = Im P_o $. 
One sees that $\, Ker P_o$ is such that $\, V= W \bigoplus Ker P_o $, 
allowing a proof of prop. 1
by induction on $dim V$.

We shall use the following group theoretical setup: 
$\{C_A\}_{A=1,...,P}$ the
set of conjugacy classes of $G$,
 $|C_A|$ the number of elements in $C_A$ and
$\Gamma=\{g_A\}_{A=1,...,P}$ a system of representatives
 of these classes.
For any $A ,\ p_A $ will be the order of $g_A$, 
 $N_A = \{h\in G / hg_A h^{-1}=g_A \}$ the centralizer of 
 $g_A$ , $|N_A|$
its number of elements equal to $|G|/|C_A|$, 
$ \chi_A =\{x_{A,j}\}_{j=1,...,|C_A|} $ a system of representatives of 
$G/N_A $ ($ e$ being one of the $x_{A,j}$), 
$ <g_A> = \{e, g_A ,...,g_A^{p_A-1} \}$ the cyclic subgroup generated
by $g_A$, $ {\cal{H}}_A = \{h_{A,a}\}_{a=1,...,|N_A|/p_A } $ a system of 
representatives of $ N_A /<g_A> $.    
                      
Equivalently, for any $g\in G$ there exist a unique couple $(A,j)$ such that 
$g = x_{A,j} g_A x_{A,j}^{-1} $ and
 $p_A = inf\{n/ g^n =e \}$.
$A$ being fixed,
 any $x\in G$ can be written uniquely 
$ x= x_{A,j} h \ ,\ x_{A,j}\in \chi_A\  ,\ h\in N_A $  
and any $h\in N_A $  can be written uniquely 
\vskip .3cm 
$h=g_A^k h_{A,a} = h_{A,a} g_A^k \ , $
 $ \ k\in \{0,...,p_A-1\}\ ,\       h_{A,a}\in {\cal{H}}_A $.
When working within a given class $C_A $ we will denote $x_{A,j}$ and
$h_{A,a}$ simply by $x_j $ and $ h_a $.

      \vskip .3cm 

{\bf Proposition 2}: the left regular representation of $\D$ is the direct sum
of $|G|$ representations $(W_g)_{g\in G}$\  : 
 $\D = \bigoplus_{g\in G} W_g, $  \ where \\  
\vskip .1cm
$ W_g = Span\{  \elt{k}{x} /x^{-1}kx = g \}$; $dim(W_g)= |G|$. \\  
If $g$ and $g'$ are  
conjugate, $W_g$ and $W_{g'}$ are equivalent representations.
\vskip .1cm
Proof: $W_g$ is the image of $\D$ by the projection  
$ a \longrightarrow a\cdot \varepsilon_g $ where the elements  
$\varepsilon_g = \elt{g}{e} $ form a decomposition of $1_{\D} $ into
a sum of orthogonal idempotents.\\
Denote a given $ g_o \in G \ \ \ g_o = x_og_A x_o^{-1} $ where $x_o$ is 
one of the $x_{A,j}$ ($x_o=e$ if $g_o =g_A$);
 then a basis of $W_{g_o}$ is
$ \left( \elt{x_j g_A x_j^{-1} }{x_j h x_o^{-1} } 
   \right)_{ (h,x_j)\in N_A \times \chi_A } .           $ 
An intertwiner $\Phi_{g_A}^{g_o} $  from $W_{g_A}$ to $W_{g_o}$ is
$$ \Phi_{g_A}^{g_o} \left( \elt{x_j g_A x_j^{-1}}{x_j h} \right) = 
   \theta_{x_j g_A x_j^{-1} }   ( x_j h ,\ x_o^{-1}) \ 
              \elt{x_j g_A x_j^{-1}}{x_j h x_o^{-1} }\  \  $$
The commutativity of $\Phi_{g_A}^{g_o} $ with left multiplication by
$\elt{g_1}{x_1}$ results from identity (\ref{a1}) below with
$(g,x,y,z)=(g_1 ,x_1 ,x_j ,x_o^{-1})$. 
  \vskip .3cm 
{\bf Proposition 3}: Each $W_{g_o}  (g_o=x_og_A x_o^{-1} )$ 
splits into $p_A$ subrepresentations
$(W_{g_o ,\l_{A,i} })_{i=1,...,p_A}$ which are eigenspaces for  the 
action of the central element $\vm = \sum_{k\in G} \elt{k}{k}$
with eigenvalues $\l_{A,i}$ (denoted in short $\l_i$ ) which are the
$p_A$th-roots of ${\o}_A = \prod_{n=0}^{p_A-1} {\o}(g_A,g_A^n,g_A) $. 
For
$(g,y)\in G^2 $ such that $y^{-1}g y =g_o$, set
\beq \psi_{\l_i,g,y}=\sum_{k=0}^{p_A-1} \l_i^{p_A-1-k}
          \prod_{n=0}^{k-1} {\o}(g,g^n y,y^{-1}gy) \elt{g}{g^k y}. \eeq
\bea \hbox{Then} \ \ \qquad  \ \ \vm \cdot \psi_{\l_i,g,y} &=& \l_i 
                                   \psi_{\l_i,g,y}              \\
                \hbox{and}\ \ \qquad \ \ \psi_{\l_i,g,gy} &=& 
\frac{\l_i}{{\o}(g,y,y^{-1}gy) } \psi_{\l_i ,g,y} ,                 \eea
so that a basis of $W_{g_o,\l_i}$ is 
$(\psi_{\l_i ,x_j g_A x_j^{-1},x_j h_a x_o^{-1} } )_{(x_j ,h_a)
                                 \in\chi_A\times {\cal H}_A }  $ .\\   

Proof: For any $\elt{g}{y}\in W_{g_o}$, and for any  
 $k\in \{0,...,p_A-1\} $,  set : 
$\phi_k = \prod_{n=0}^{k-1} {\o}(g,g^n y,y^{-1} gy) \elt{g}{g^k y}$.

The characteristic polynomial of the action of $\vm $ on the space 
spanned by the $\phi_k$'s is then (cf (\ref{a4}) :
$$ det(\l -\rho_{reg}(\vm ) ) = \, \, 
\l^{p_A} - \prod_{n=0}^{p_A -1} {\o}(g,g^n y,y^{-1} gy) = \l^{p_A} -{\o}_A ,$$
leading to  eigenvectors $\psi_{\l_i ,g,y} $ 
 which, altogether, are a generating
system of $\, W_{g_o ,\l_i }$. Each eigenspace $W_{g_o, \l_i}$ 
is stable under 
the action of $\D$ because $\vm$ is central 
and the $\l_i$'s are distinct.\\ 

{\bf Definition} : $A$ being fixed let $(\pi,V)$ be a projective representation
of $N_A$ on a vector space  $V$, with cocycle $\theta_{g_A}$, one 
has in $End(V)$ :
\beq \pi(h_3) \pi(h_2)\,  = \, \, 
\theta_{g_A}(h_3 ,h_2 ) \pi(h_3 h_2)\ . \label{proj} \eeq
Then, following~\cite{dpr}, we define the "DPR-induced" of 
$(\pi , V)$ as the representation
 $(\rho_{\pi}, C [\chi_A ]\otimes V)  $
of $\D$ given by:
\beq \rho_{\pi}(\elt{g}{x}) |x_j>\otimes|v> =
\delta_{g,x_k g_A x_k^{-1} } \ 
\frac{\theta_g(x,x_j)}{\theta_g(x_k,h_2)} |x_k>
\otimes \  \pi(h_2)|v>                             \label{rhopi}
 \eeq
where $|v>\in V $ and  $(x_k,h_2)\in \chi_A\times N_A $ are defined by
$\, x\,  x_j\, =\, x_k \, h_2$.\\ 

\vskip .3cm 
{\bf Proposition 4}: For any $g_o\in C_A , \  (\rho_{reg},W_{g_o}) $ 
is equivalent to the representation DPR-induced from 
$(\pi_{g_o}, C[N_A])$ given by : 
$$ \pi_{g_o}(h_2)|h_1> = \theta_{g_A}(h_2, hx_o^{-1}) |h_2 h>       $$
Proof: An explicit intertwiner 
$\Theta_{g_o}: W_{g_o} \longrightarrow \C[\chi_A]\otimes \C[N_A]   $ is:  
\beq \Theta_{g_o}(\elt{x_j g x_j^{-1}}{x_j h x_o^{-1}} )=
\frac{1}{\theta_{x_j gx_j^{-1}}(x_j ,hx_o^{-1}) }  \   
     |x_j>\otimes |h> .\label{INTTHETA} \eeq
Also note  that $\pi_{g_o}$ is equivalent
 to $\pi_{g_A}$ with intertwiner
\\  $\Psi :  (\pi_{g_A}, \C[N_A])\longrightarrow (\pi_{g_o}, \C[N_A])$,
$\Psi |h> = \theta_{g_A}(h,x_o^{-1}) |h>$.

Having these equivalences of representations we can now focus 
on the subrepresentations of the $\, W_{g_A}$'s :\\ 
 \vskip .3cm 
{\bf Proposition 5}: $(\rho_{reg}, W_{g_A,\l_i}) $ is equivalent to the 
representation DPR-induced from $(\pi^{\l_i}, \C[{\cal H}_A] )$ where :
\beq \pi^{\l_i}(h_2)|h_a> =
 \frac{\l_i^l}{\prod_{n=0}^{l-1} {\o}(g_A,g_A^n ,g_A)} \ 
 \frac{\theta_{g_A}(h_2,h_a)} {\theta_{g_A}(h_b,g_A^l) } \  |h_b> 
                                          \label{pilambda}      \eeq  
$(l,h_b)\in \{0,...,p_A-1\} \times {\cal H}_A $ being defined by
$h_2 h_a = g_A^l h_b \ .$\\ 
\vskip .1cm
Proof: The fact that $\pi^{\l_i}$ satisfies (\ref{proj})
 results from (\ref{a5}).
\beq \Omega (\psi_{\l_i ,x_j g_A x_j^{-1} ,x_j h_a } ) =
     \theta_{x_j g_A x_j^{-1} }(x_j ,h_a) \  
     \Theta_{g_A}(\psi_{\l_i,x_j g_A x_j^{-1}, x_j h_a} )
        \label{INTOMEGA} \eeq
is an intertwiner 
$W_{g_A,\l_i}\longrightarrow \C [\chi_A]\otimes \C [{\cal H}_A] $,
as a rather tedious computation shows.
 \vskip .3cm 
{\bf Proposition 6:} $(\pi_{g_A}, \C[N_{g_A}])$ is the direct sum of
$p_A\ \  \theta_{g_A}-$projective 
representations $N_{A,\l_i} $ equivalent
to the $(\pi^{\l_i}, \C[{\cal H}_A] )$,  
$\C[N_{g_A}]= \bigoplus_{i=0}^{p_A-1} N_{A,\l_i} $.
\vskip .1cm
Proof: 
\beq \Pi |h_a> = \sum_{m=0}^{p_A-1} \theta_{g_A}(h_a,g_A^m)
   \frac{  \prod_{n=0}^{m-1} {\o}(g_A,g_A^n,g_A) }{\l_i^m}
   \     |g_A^m h_a >                            \label{Pi}  \eeq
is an intertwiner 
$(\pi^{\l_i}, \C[{\cal H}_A]) \rightarrow (\pi_{g_A}, \C[N_{g_A}]) $
whose image we call $N_{A, \l_i}$.
\beq {\cal P} |g_A^k h_a> = 
  \frac{\l_i^k}{\prod_{n=0}^{k-1} {\o}(g_A, g_A^n, g_A) 
                  \theta_{g_A}(h_a, g_A^k)  }  
                                      \    |h_a> \label{calP} \eeq
is an intertwiner 
$(\pi_{g_A}, \C[N_{g_A}]) \rightarrow (\pi^{\l_i}, \C[{\cal H}_A] )$. 
$ \Pi \circ {\cal P} $ gives an explicit expression for the projector
$\C[N_{g_A} ] \rightarrow N_{A,\l_i} $ .\\

We now reach our main point: \\   
\vskip .3cm 
{\bf Proposition 7 :} The irreducible representations of $\D$ are
(up to equivalence) characterized by a couple $(A, \alpha )$ where
$A$ is the label of a conjugacy class in $G$ and $(\alpha , V_{\alpha})$
is an irreducible projective representation of $N_{g_A}$ with 2-cocycle
$\theta_{g_A} $ which can be realized as a subrepresentation of 
$ (\pi_{g_A}, \C[N_{g_A}] )$. Furthermore 
$\alpha(g_A)\, = \, \l_i \, Id_{V_{\alpha} }$, for one $\l_i$, so that 
$(\alpha, V_{\alpha} )$ is equivalent to a subrepresentation of
$N_{A, \l_i} (\sim \pi^{\l_i} )$. Then the corresponding representation of 
$\D $ is equivalent to the DPR-induced  
$(\rho_{\alpha} , V_{A \alpha }=\C [\chi_A ]\otimes V_{\alpha} )$.\\ 
\vskip .1cm 
Proof: This is clear once one has noticed that:
$$ \pi^{\l_i}(g_A) = \, \l_i \ 
  \frac{\theta_{g_A}(g_A, h_a) }{\theta_{g_A}(h_a, g_A) } 
  Id_{\C [{\cal H}_A ] } \, = \, \l_i \,  Id_{\C [{\cal H}_A ] }        $$
(so that $\pi^{\l_i}\, $ is  
$\, \theta_{g_A}$-projective, and $g_A^{p_A}=e$ 
implies $\l_i^{p_A}={\o}_A $ ) and 
\bea dim \D  &=& \sum_A |C_A| dim W_{g_A} 
    = \sum_A |C_A| |\chi_A |   dim \C[N_{g_A}]      \\     
    &=& \sum_A |\chi_A|^2 \sum_{\alpha} dim^2 (V_{\alpha}) = 
        \sum_{A \alpha} dim^2 (V_{A \alpha }) \  .   \eea  

  From (\ref{rhopi}) one gets:\\ 
  \vskip .3cm 
{\bf Proposition 8 :} The characters of the representations of $\D $ are:
\beq \chi_{A,\alpha}(\elt{g}{x} )= \delta_{g\in C_A}  
     \delta_{x\in N_g} \  
\frac{\theta_g(x,x_j)}{\theta_g(x_j,h) }  
                    \ \chi_{\alpha}(h)  \label{car}
\eeq
where $N_g$ is the centralizer of $g$, and 
$(x_j, h)\in \chi_A \times N_{g_A} $ are defined by 
$g\, = \,x_j\, g_A \, x_j^{-1},$   $\, x\,= \,x_j\, h \,x_j^{-1}$.

\vskip .3cm 
{\bf Summary of the results.}

%
These eight propositions, which concatenated 
in Spinoza's style some partly available results,   
 first proved the semi simplicity of the twisted 
double algebra. They also described the
 irreducible representations 
 from the regular representation and used an interesting
 generalization of character theory for quasi-Hopf algebras.

\vskip .5cm 
{\bf {\large Perspectives  } }
\vskip .2cm 
\par These results allow the computation of
 the Verlinde $S$ matrix
 using the ribbon functor applied to the coloured Hopf link. 
 The complete formulae, involving a number of phases 
  (i.e. exponentials of the additive 3-cocycle)are presented 
   in~\cite{costegr2000}. These phases are absolutely non trivial 
  in the general case of any non abelian group and in the
   so-called "non cohomologically trivial" situation.\\  
   
   It is worth noticing that such rather complicated objects appear,
 if one wants to classify 
   conformal theories. Classifying 
   such theories with a small number of primary fields will use
   classification of {\em finite groups} (and computing their 3-cocycles)
 with a small 
   number of conjugacy classes, a task performed, up
 thirteen classes, by~\cite{VL}. 
It has been noticed, 
   at least since E. Landau that there are only a finite number of 
finite groups 
   with a given number of classes (but how many ? may be physicists inspired 
   by recent classifications or statistical approaches could give 
interesting comments 
   to this question). Then, there seems to be many more 
modular invariant partition 
   functions than in the affine case. \\        
   
   These 3-cocycles  have been considered in the context of 
string and membranes 
    studies using  names such that 
"orbifolds with finite torsion". One idea is  
    that any CFT can live on a string worldsheet, and our algebra 
    encodes twisted boundary conditions, in the sense
 of G. t'Hooft and C.P. Korthals-Altes, which is known to relate to 
non-commutative geometry. 

      \vskip .5cm
\vskip .5cm
{\bf Acknowledgements:}  One of us (A.C) 
 would like to thank P. Bantay, T. Gannon, P. Roche, P. Ruelle, 
V. Pasquier, J. B. Zuber 
for interesting discussions, and also J. P. Bourguignon and IHES 
 for kind hospitality.
\vskip 1.5cm
\vskip 1.5cm
\pagebreak 

\appendix{\bf Appendix : Group cohomology identities}
\vskip .1cm 
$  \theta_g \, $ defined above satisfies:  
\beq \theta_g(x,y) \  \theta_g(xy,z)\,  =\,\,  \theta_g(x,yz) 
                   \  \theta_{x^{-1}g x}(y,z) \label{a1}  
    \eeq
Proof: For instance simplify successively (\ref{a1}) with the help of
(1) written with $(g_i)_{i=1,...,4} = (x,y,z,(xyz)^{-1}gxyz); $
$ (x,y,(xy)^{-1}gxy,z); $ $(g,x,y,z)$ and finally 
$(x,x^{-1}gx,y,z). $ 
\beq \hbox{For any}\  g,y \ \ \ \  
\prod_{n=0}^{k-1} {\o}(g,g^n y,y^{-1} gy) = \  \  
     \theta_g (g^k ,y) \prod_{n=0}^{k-1} {\o}(g,g^n ,g) \label{a2} 
\eeq        
Proof: Let us form the product of the $\, k$ identities
 obtained from (1) \\  
with
$\, (g_i)_{i=1,...,4} \, = \, (g,g^n ,y,\ym gy)\,  $ for $\, n=\, 
0,...,k-1$,  and divide it by
 a similar product for $\, (g_i)\, = \,\, (g,g^n ,y,\ym gy)\, $ :
%
\beq 
   \prod_{n=0}^{k-1} {\o}(g,g^n y,y^{-1}gy) =
   \theta_g(y,y^{-1} g^k y) \  
    \prod_{n=0}^{k-1} {\o}(y^{-1}gy,y^{-1}g^ny,y^{-1}gy)   \label{a3}  \eeq
Proof: Let us build similar products of identities for $$(g_i) =  
(g,y,\ym g^n y,\ym g y), \ (\ym ,gy,\ym g^n y,\ym gy), \ 
(\ym ,y,\ym g^n y,\ym g y)                     $$ and simplify a factor
involving five ${\o}$'s by the identity written with 
$(g_i)=(\ym ,y,\ym gy, \ym g^k y) $.

When $k=p_A $ and $g\in C_A \ (g^{p_A}=e)  $ 
 (\ref{a2}) and (\ref{a3}) insure that
\beq \prod_{n=0}^{p_A-1}  {\o}(g,g^n y,\ym g y) =
     \prod_{n=0}^{p_A -1} {\o}(g,g^n,g) = 
     \prod_{n=0}^{p_A -1} {\o}(g_A, g_A^n ,g_A) ={\o}_A  \label{a4}  \eeq

For any positive integers $k,l$ and any $ g\in G $:
\beq \prod_{n=0}^{k+l-1} {\o}(g,g^n,g) = \theta_g(g^l,g^k) 
\prod_{n=0}^{k-1} {\o}(g,g^n ,g) \prod_{n=0}^{l-1} {\o}(g,g^n ,g) \label{a5}\eeq
\beq \hbox{and this implies the symmetry :  }\ \  
 \theta_g(g^k ,g^l) =  \theta_g(g^l,g^k) .              
                                            \label{a6}  \eeq 
Proof: Set
 $$Q_k = 
 \frac{ \theta_g(g^l,g^k)  }{ \prod_{n=k}^{k+l-1} {\o}(g,g^n,g) } \  ,  $$
   (1) with 
$(g_i) = (g,g^l,g^k,g); (g^l,g,g^k,g) $ implies that $Q_{k+1}=Q_{k}$, 
therefore equal to $Q_0 =1/ \prod_{n=0}^{l-1} {\o}(g,g^n,g) $.
            
Note for completeness the corresponding identity with negative powers: 
\beq \prod_{n=0}^{k+l-1} {\o}(g,g^{-n},g)\,  = \,\,  
     \frac{{\o}(g,g^{-l},g)}{\theta_g(g^{-l},g^{1-k}) } \ 
     \prod_{n=0}^{k-1} {\o}(g,g^{-n},g)
     \prod_{n=0}^{l-1} {\o}(g,g^{-n},g)                   \label{a7} \eeq
 and the relation between positive and negative powers :
\beq \prod_{n=0}^l {\o}(g,g^{-n},g) = \theta_g(g^{-l},g^l) 
     \prod_{n=0}^{l-1} {\o}(g,g^n,g)                     \label{a8} \eeq
Proof: by induction and use of (1) with $(g_i) =
       (g,g^{-l-1},g^{l+1},g)$ $ (g,g^{-l-1},g,g^{l+1})$ 
     $ (g^{-l},g,g^l,g) (g,g^{-l},g^{l},g) $.

\vskip .3cm 

\vskip 1.5cm
\vskip 1.5cm
\vskip 1.5cm

\end{document}